# Work-From-Home is Here to Stay: Call for Flexibility in Post-Pandemic Work Policies

Darja Smite[1,2], Nils Brede Moe[2,1], Jarle Hildrum[3], Javier Gonzalez Huerta[1], Daniel Mendez[1,4]
[1] Blekinge Institute of Technology, [2] SINTEF, [3] Telenor, [4] fortiss

**Abstract:** In early 2020, the Covid-19 pandemic forced employees in tech companies worldwide to abruptly transition from working in offices to working from their homes. During two years of predominantly working from home, employees and managers alike formed expectations about what post-pandemic working life should look like. Many companies are currently experimenting with new work policies that balance both employee- and manager expectations to where, when and how work should be done in the future. In this article, we gather experiences of the new trend of remote work from 17 companies and their sites, covering 12 countries. We share the results of corporate surveys of employee preferences for working from home and analyse new work policies. Our results are threefold. First, through the new work policies all companies are formally giving more flexibility to the employees with regards to working time and work location. Second, there is a great variation in how much flexibility the companies are willing to yield to the employees. The variation is related both to industry type, size of the companies, and company culture. Third, we document a change in the psychological contract between employees and managers, where the option of working from home is converted from an exclusive perk that managers could choose to give to the few, to a core privilege that all employees feel they are entitled to. Finally, there are indications that as the companies learn and solicit feedback regarding the efficiency of the chosen strategies, we will see further developments and changes of the work policies with respect to how much flexibility to work whenever and from wherever they grant. Through these findings, the paper contributes to a growing literature about the new trends emerging from the pandemic in tech companies and spells out practical implications onwards.

# Introduction

In March 2020, employees of most if not all tech companies in the world forced their employees to work from home in response to the COVID-19 pandemic, marking the turn of the history in the magnitude of experience and perception of work from home (WFH). Almost two years later, with the ease of restrictions and periodic attempts to reopen the society we see that the pandemic has left a permanent mark in the fundamental principles of the workplace as many information workers express their preferences to continue working from home. It's, thus, fair to assume that WFH is here to stay.

A large survey of over 30'000 American workers identified five reasons for the large shift in favour of WFH (Barrero et al., 2021 a), including better-than-expected WFH experiences, new investments in physical and human capital that enable WFH, the change in attitude and stigmatisation of remote workers, lingering concerns about crowds and contagion risks, and a pandemic-driven surge in technological innovations that support WFH. And although we know that fully remote work from home is not challenge free and is not for everybody (Ford et al., 2020; Ralph et al., 2020), there is a growing realisation that the old, good days, in which the notion of "being employed" was strongly associated with working in the office, will change. An increasing demand for flexibility from the new hires during job interviews calls for new corporate policies regarding working from home. In fact, the degree of flexibility might now become the make-or-break point in many employment decisions. The following dialogues between developers (D) and managers (M) reflect the actual stories we have heard from our industry partners in the past few months, and they show that the rules for retaining and attracting employees are indeed changing.

*Somewhere in Sweden*
M: Now it's time to return to the office.
D: No, we will continue working from home.
M: You must return.
D: OK, then we quit.
M: Please, don't. You can continue working from home.

*Somewhere in Norway*
D: May I work from home; I broke my leg.
M: Sorry, we just introduced a new policy. You are to be in the office 3 days a week.
D: But my situation is different!?
M: Sorry, but the rules are equal for everybody.
D (thinking): Oh, maybe I should start looking for a new job…

*Somewhere in Germany*
D: Since I am working from home during the pandemic, may I join my family in India and work from there?
M: OK, you are allowed to do that, but we will cut your salary to the level of the salary in India, since your living standard is going to be lower.
D: (thinking): Oh, maybe I should start looking for a new job…

*Somewhere in Brazil*
D: Can I get a raise?
M: Not really. It's a tough time now.
D: OK, then I quit.
M: Why? Aren't you satisfied to work for us?
D: I was, but then I got an offer from a start-up in the US that is ready to pay a much higher salary in US dollars.

On the other hand, there is a growing interest in understanding the long-term effects of isolation and remote work. A Microsoft study with over 60'000 employees shows that firm-wide remote work made the collaboration network more static with fewer new ties being established and heavily siloed, with fewer ties that cut across formal business units, as communication shifted from synchronous to asynchronous (Yang et al,, 2021). Based on related research that suggests that changes in collaboration and communication media impede knowledge transfer and the ability to convey and process complex information, as well as reduce the quality, the authors further expect these network changes to have a negative impact on productivity and innovation (Yang et al,, 2021). Other researchers have expressed similar concerns regarding the long-term effects of deteriorating social ties (Clear, 2021) and decreased interest in collaborative work when doing it remotely (Kane et al. 2021; Smite et al., 2021).

One key approach managers are using to tackle these problems is formulating new work policies that set the expectation to employees about both office presence and behaviours in a new hybrid work situation. While the existence of such policies is nothing new, the fast-growing adoption and variation of such policies during the pandemic is huge (Choudury, 2020). In this article, we portray the different trends and the variety of options incorporated in the new work policies of 17 companies and discuss how these companies have adjusted their strategies to address the demand of the new reality.

# Overview of the cases and data

The data behind our findings comes from 17 companies that differ in domain and size, from small one-office companies to large international companies represented by several offices (see an overview of our dataset in Table 1). In this article, we present the results of 23 company-internal surveys of employee preferences for working from home or in the office (see the summary in Table 1), and 26 corporate post-pandemic work policies (see the summary in Table 2).

**Surveys:** Employee preferences for WFH were elicited through corporate surveys designed in the companies. We had access to the results of 23 surveys from 22 corporate entities (including 2 runs of the survey in one company). The summary in Figure 1 portrays the voice of 11'318 respondents. The response rates in different companies varied from the min of 15% to the max of 100% with 65% as the median. Since the scales of possible response options across surveys varied, we integrated the results using a common scale of frequency of WFH: "Never" (meaning only in the office), "Occasionally" (less than once a week), "Less than half week" (1-2 days a week), "Half the time", "More than half week" (3-4 days a week), "Occasionally in the office" (less than once a week in the office), and "Always" (see Appendix A for the individual survey scales). Notably, not all response options have been available in each of the surveys (for example, some surveys offered respondents to choose "50:50%", while others had 1 day/week, 2 days/week, etc.). This is why, we recommend looking at the three major trends with respect to the amount of responses in favour of primarily office work ("Never" and "Occasionally"), those in favour of primarily remote work ("Always" and "Occasionally in the office"), and those who give preference to flexible or hybrid work arrangements ("Less than half week", "Half the time" and "More than half week").

Further detailed analysis was performed in selected cases. First, in two companies, we yielded to see whether respondents change their preferences for WFH over time (presented in Figure 2). We also received results of stratified analysis of role and age in the Storebrand's survey. Stratified analysis was also performed for the survey from Telenor Norway, using ANOVA factor analysis (Allen, 2017), for the preferences for WFH, including gender, role, age and commute time (presented in Figure 3). We have used n-way ANOVA analyses to identify whether preferences for WFH might be dependent on different factors or their interactions and performed Tukey *post-hoc* tests to identify which groups in a particular factor have different preferences. We only report differences on preferences in the stratified analysis when they have been found to be statistically significant ($p < 0.05$ with $\alpha = 0.05$). In addition, we ran a set of logistic regressions (Menard 2002) using as the dependent variables binary measures of employees' stated preferences to work from home and the same explanatory variables. Based on these regressions, we present the probabilities to work from the office and the related marginal effects of each variable.

**Work policies:** Next, we reviewed post-pandemic work policy documents. We have gathered 26 policies, some designed to regulate the WFH in a single office, others represent centralised corporate efforts. We analysed the options permitted by the corporate policies (summarised in Figure 4), and the exceptions from the general rules, if any. Corporate policies were discussed with representatives from each of the companies in informal and/or formal interviews. During these interviews, we additionally inquired whether it is permitted to move within the country or globally, and how the companies support home office equipment of those working remotely.

**Table 1:** Overview of the data collection activities and collected data

| Corporate cases included in the study | Case ID | Locations | Size | Survey of WFH preferences | | | WFH policy, status |
|---|---|---|---|---|---|---|---|
| | | | | N | Rate | Executed | |
| 'InterSoft' (pseudonym): Large international company delivering music streaming services. R&D locations. | A | Sweden, US, the UK | 1'411 | 1408 | 100% | Mar-Apr 2021 | "Work from anywhere", Fall 2021 |
| 'GlobCo' (pseudonym): Large multinational company delivering software intensive systems for the telecom market. A Swedish site and several Indian sites in one business unit and a Chinese site in another business unit. | BA | Sweden | 358 | 276 | 77% | Mar 2021 | "Instruction Remote Work in Sweden", Fall 2021 |
| | BB | India | 961 | 769 | 80% | Mar 2021 | "Flexible work guidelines", Dec 2021 |
| | BC | China | 54 | 35 | 65% | Mar 2021 | "Alignment of practice", Dec 2021 |
| **Malvacom**: Small consultancy company delivering applications and server solutions. Whole company | C | Sweden | 17 | 13 | 75% | Nov 2021 | Policy for remote work, Oct 2021 |
| **Storebrand**: A large software company leading the Nordic market in long-term savings and insurance. Whole company. | D | Norway | 1'300 | 358 | 26% | Jan 2022 | "Future Storebrand", Spring 2021 |
| **SB1 Utvikling**: Large software development company owned by a Norwegian alliance of banks. Whole company. | E | Norway | 355 | 230 | 65% | Apr-May 2021 | "Create rule of conduct together", Jun 2021 |
| | | | | 224 | 63% | Sep-Oct 2021 | |
| **Sbanken**: Scandinavian online-only bank offering financial services. Whole company. | F | Norway | 344 | 209 | 61% | Nov 2021 | "Sbanken post COVID", Fall 2021 |
| **KnowIT**: Large consultancy company delivering digital transformation and systems development services. One Norwegian site. | G | Norway | 175 | 138 | 79% | Sep-Oct 2021 | "Guidelines and recommendations for hybrid workday", Fall 2021 |
| **Kantega**: Medium-size software company developing bespoke software for diverse markets, incl. financial and public sector. Whole company. | H | Norway | 172 | 162 | 94% | Jun 2021 | Principles for future workplace, Fall 2021 |
| **Blank**: A small employee-owned consultancy company offering software development services and developing own products. Whole company. | I | Norway | 44 | 36 | 82% | Feb 2021 | Flexible work principles (established with the company foundation) |
| | | | | 40 | 91% | Dec 2021 | |
| 'KNor' (pseudonym): A large Norwegian company developing embedded software products. One development department. | J | Norway | 3'676 | 395 | 58% | May 2020 | "Guideline for Hybrid Working", Jan 2022 |
| **Tietoevry**: Large multinational technology company delivering software solutions to customers worldwide. Detailed analysis of five largest geographic locations and a corporate survey from 16 geographic locations. | KA | Norway | 3'881 | 1026 | 28% | Apr-May 2021 | Norwegian hybrid working guidelines, Oct 2021 |
| | KB | Sweden | 3'270 | 1102 | 28% | | Swedish hybrid working guidelines, Oct 2021 |
| | KC | Finland | 2'954 | 793 | 24% | | Finish hybrid working guidelines, Oct 2021 |
| | KD | India | 2'542 | 593 | 20% | | Indian hybrid working guidelines, Nov 2021 |
| | KE | Czech Rep | 3'676 | 390 | 15% | | Czech hybrid working guidelines, Nov 2021 |
| **Telenor**: Large multinational market leading telecom operator. Four business units. | LA | Norway | 3'446 | 2219 | 65% | 2021 | Guidelines flexibility in 'TNor' Group, Aug 2021 |
| | LB | Sweden | 1554 | – | – | – | Rules of Engagement, Jul 2021 |
| | LC | Denmark | 1'086 | – | – | – | "Flexible work", Jun 2021 |
| | LD | Finland | 1'487 | – | – | – | "Flexible work model", Nov 2021 |
| 'FSwed' (pseudonym): Large software development company delivering financial systems and services. Whole company. | M | Sweden | 589 | 429 | 73% | Mar-Apr 2021 | WFH update on intranet, Oct 2021 |
| 'SpanCo' (pseudonym): Part of a large consultancy company developing software for the banking sector. A Spanish site. | N | Spain | 13 | 13 | 100% | | Employment contract, Sep 2021 |
| **GFT**: Medium-size European consultancy company delivering software solutions for finance and insurance sectors. Detailed analysis from one site, and survey data from five Spanish sites. | O | Spain | 1'800 | 1045 | 72% | Spring 2021 | Policy for remote work, Oct 2018 /2020 |
| **CQSE**: Small product development and consultancy company specialised in software maintenance and evolution. A German site. | P | Germany and USA | 50 | – | – | – | Flexible remote work policy, prior to COVID-19 |
| **QualityMinds**: A medium-sized consultancy company focusing on software quality, quality assurance and development. Whole company. | Q | Germany and Poland | 250 | 90 | 36% | Oct 2021 | "QM Company Guide", 2013 |

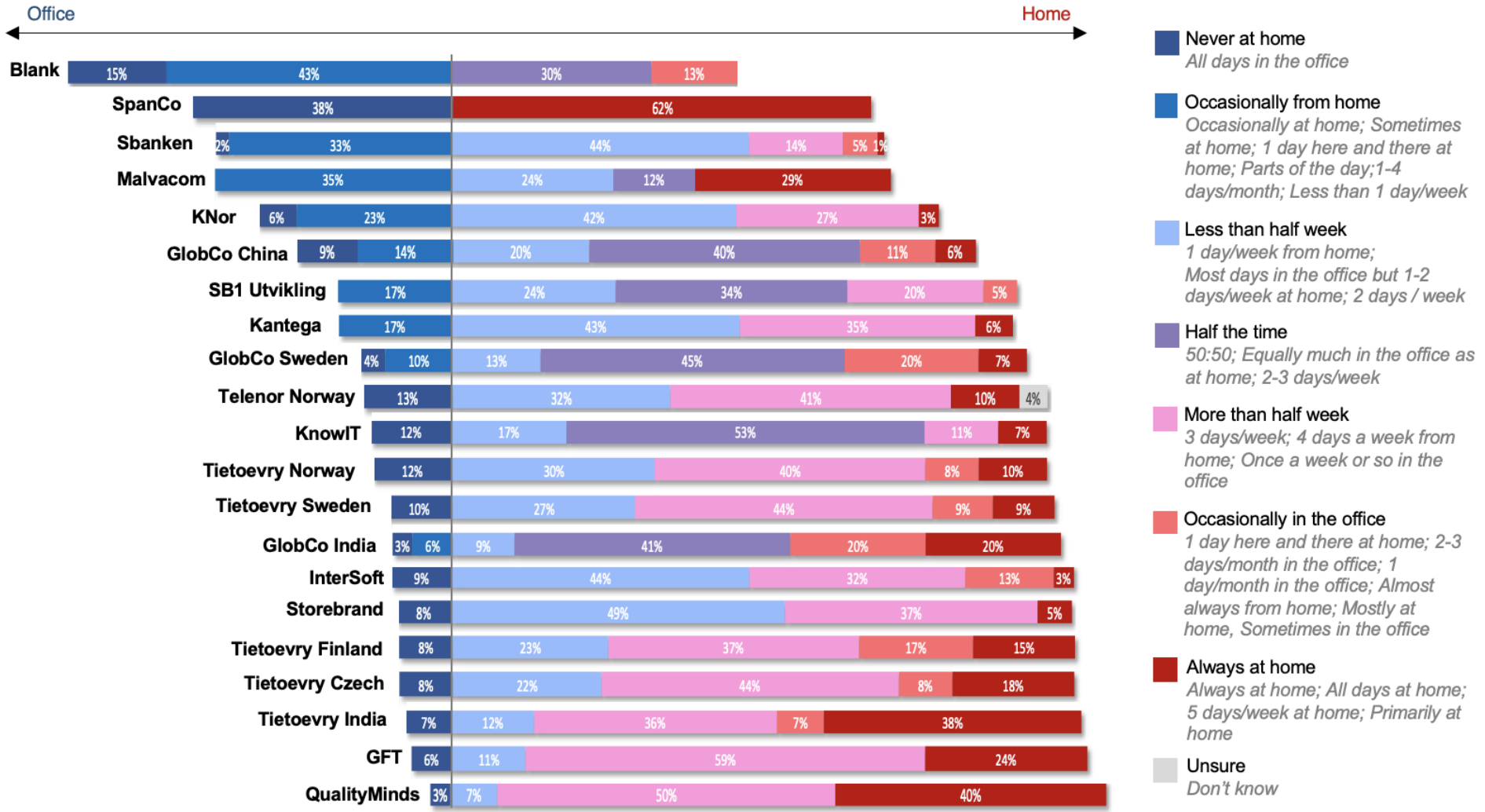

Figure 1: Overview of the employee votes for working from home options.

# Employee preferences: How much do employees want to work from home?

To understand the future demand for flexibility, we integrated the results of corporate surveys of employee preferences for working from home (see Figure 1). What did we find?

**How many want to return to the office?**
Surprisingly, there are not that many employees who will return to the traditional way of working "at the office 9 to 5 five days a week" (the median value for employees who opted for "Never at home" and "Occasionally from home" options is only 10%). Groups of employees who choose to work only from the office reach the maximum of 38% in the special case of the Spanish SpanCo, where the employees have been given only the binary choice between full time office work and full time remote work, with the Norwegian Blank on the second with 15%.

**How many do not want to return to the office at all?**
Groups who want to only work from home ("Always" category) vary from 0% in the Swedish Malvacom, the Norwegian SB1 Utvikling and Kantega, to 38% in Indian Tietoevry, 40% in the German QualityMinds, and 62% in the exceptional case of the Spanish SpanCo, the votes of which reflects the binary choice. Interestingly, the median value of employees who opted for "Always" at home and "Occasionally in the office" mirrors the office workers – 10%. Five companies from Germany, India and Spain in our dataset report that at least every fifth employee prefers to continue to work fully remotely.

**How prevalent is the demand for flexibility?**
The vast majority of respondents in our dataset choose to mix days in the office with days at home, which recently gained a label of *a hybrid work setup*. The main motivation for this trend is that home offices are typically associated with way more superior conditions for concentration and uninterrupted work, while offices are seen as way more superior for collaborative work (Smite et al., 2022). There are, however, some exceptions to this trend. A recent survey conducted by the Gensler Institute (2021) suggests that younger generations of employees who often live in homes unsuited for concentration work are less productive working from home than from the office, and want to use the office for uninterrupted work.

The preferred proportion of time spent in the office vs home varies across companies and within companies. In fact, our data suggests that in companies that surveyed their employees several times, even the individual opinions have changed over time (see Figure 2). This emphasises that the demand for WFH is situational and motivates the high degree of flexibility.

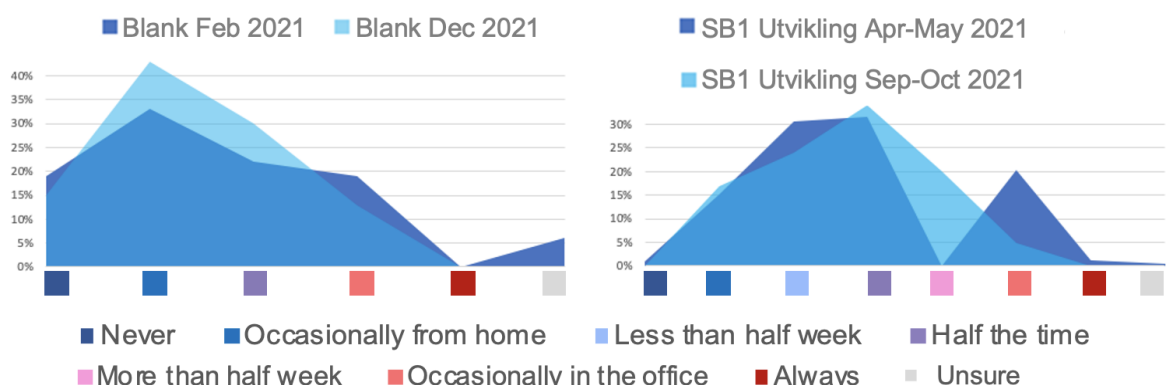

**Figure 2:** Changes in preferences for WFH overtime at Blank and SB1 Utvikling. Result: No undecided, fewer extremes, more opt for hybrid with more office presence in Blank and less in SB1 Utvikling.

Interestingly, in Tietoevry, which practiced WFH since before the pandemic, we were able to compare the *pre-pandemic working* with *the employee preferences for post-pandemic working*. To our surprise, ⅕ of previously "mostly office" employees would like to visit the office less than once a week, while roughly ⅕ of "mostly remotes" would like to go to the office at least 3-4 days/week. This means that the attitudes towards remote work have changed significantly and that companies cannot predict the employee preferences based on the historical choices.

**Who chooses which preferences?**
Wherever possible, we performed detailed analysis to check for demographic predictors of the preference for office or home working (see Figure 3). The candidate factors were age, tenure, gender, managerial role or not, size of the cities, countries and pre-pandemic choices).

*Role:* When contrasting engineers' vs managers' choices in Storebrand (see Figure 3A) and Telenor Norway (see Figure 3B), we found that managers prefer to spend more days at the office than other employees.

*Age:* In Storebrand (see Figure 3C) and Telenor Norway (see Figure 3D), we checked employee preferences for WFH across different *age groups*, and found that those younger employees (<29 in Telenor) prefer more office presence than senior employees (>56 in Telenor), who prefer more WFH days.

*Commute time* might also be one important factor when predicting the willingness of the employees to come into the office. In Telenor Norway (see Figure 3E), we analysed employee preferences in Oslo, the capital of Norway, compared with smaller Norwegian towns, assuming that the commute time in the capital city is longer. We found that the number of employees who want to work in the office is three times higher in smaller towns. Unwillingness to spend time commuting was found the No 1 reason for employees not present in the office in three other Norwegian companies – SB1 Utvikling, Storebrand and Sbanken. To exemplify the length of commute, in SB1 Utvikling 36% of employees spend 1,5 hours commuting to and from work each time they go to the office.

*Gender:* Our analysis of male and female preferences for WFH in TietoEVRY (see Figure 3F) and Telenor Norway (see Figure 3G) did not indicate any significant differences in preferences for working from home overall. However, when analysing differences within the age-group of 30-40 years in Telenor, we find a slightly but significantly higher preference to work from home for females. As many people in this age-group have small children, one potential explanation is that females choose to work from home more than the males for reasons of caring for children.

This finding is of particular interest because telework historically has been tightly associated with gender segregated motivation (Pratt, 1984). For instance, Nguyen and Armoogum (2021) found the gender-divided preferences towards WFH to vary, with higher preferences among female professionals. Such differences have also been found by Bloom et al. (2021, a) who point out that among college graduates with young children, females want to work full time from home almost 50% more than males. Although our analysis of data from Telenor Norway does not account for employees having small children, it still aligns with this finding.

Other candidate predictors that we have not managed to analyze include different *family situations* (living alone vs living with a spouse vs living with a family with kids (further divided into different age groups)), which might gravitate people towards working from home or from the office, as well as *organisational and national cultures*, which determine the level of autonomy, family organisation and gender-based role differences.

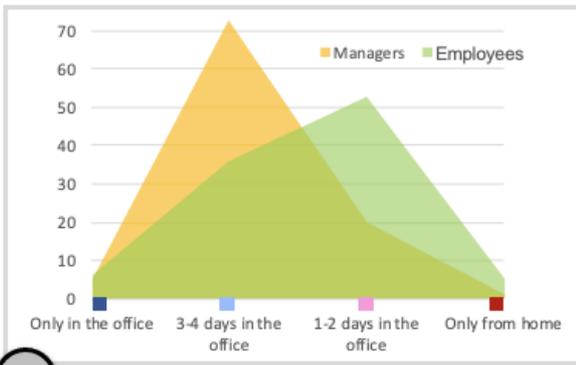

**A** by **Role** (managers vs employees) at **Storebrand** (N=358).
Result: **Managers prefer to have more office presence**

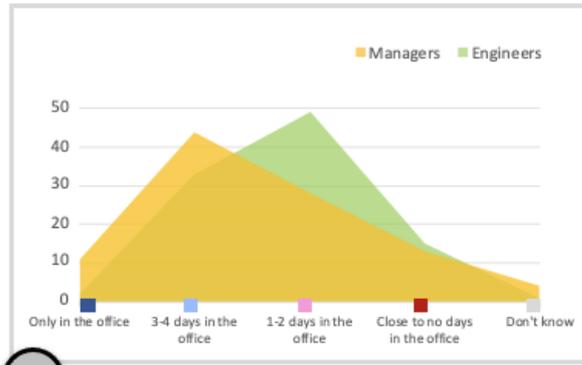

**B** by **Role** (managers vs employees) at **Telenor Norway** (N=2019).
Result: **Managers prefer to have more office presence**

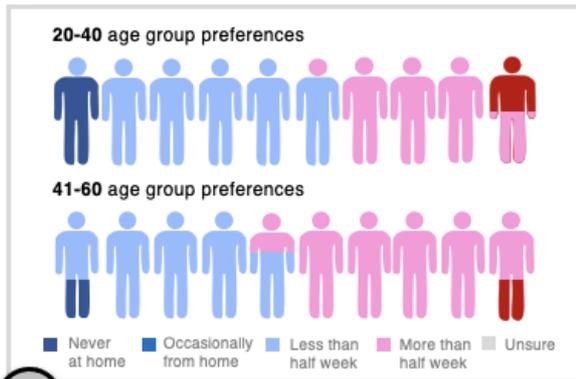

**C** by **Age** at **Storebrand** (N=358). Result: **Seniors prefer more hybrid options and slightly more work from home**

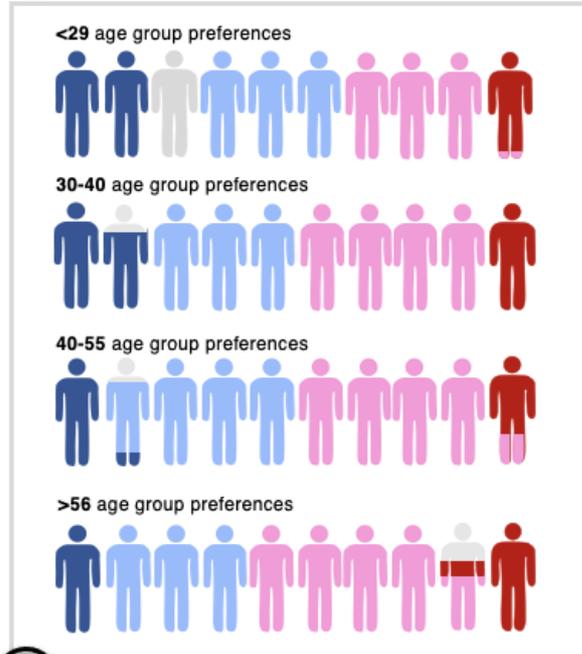

**D** by **Age** at **Telenor Norway** (N=2019). Result: **Seniority determines the willingness to work more from home, while juniority determines the preference for more office presence**

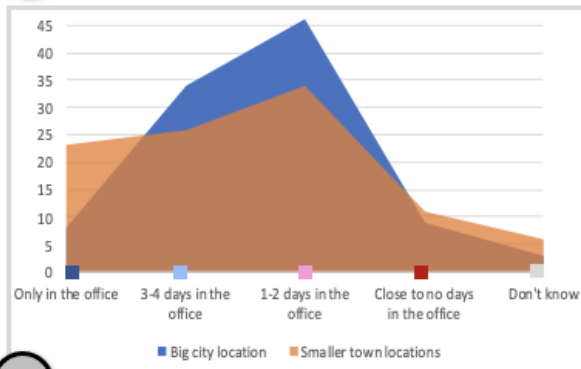

**E** by **Commute time** (Oslo vs smaller Norwegian towns) at **Telenor Norway** (N=2019). Result: **People from smaller towns choose more office presence than city inhabitants**

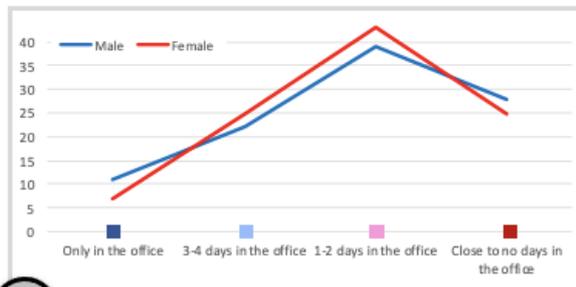

**F** by **Gender** at **Tietoevry** (N=4450 across 16 company sites). Result: **No significant differences between genders**

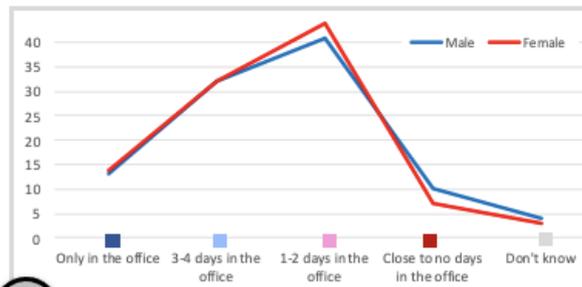

**G** by **Gender** at **Telenor Norway** (N=2019). Result: **No significant differences between genders**

**Figure 3:** WFH preference distributions in different groups of respondents.

# Post-pandemic work policies: How much do companies allow employees to work from home?

The next important question is how should companies respond to the increased needs for flexibility that we have seen in the employee surveys. Motivated to find the answer, we collected the guidelines, strategies, and policies that regulate the extent of permitted work from home from 17 companies. Figure 4 summarises our findings.

**Most companies have established new WFH policies, instructions or regulations**
Only a few companies in our study (Norwegian Blank, Spanish GFT, German CQSE and multinational Tietoevry) have long traditions of flexible work from the times prior to the pandemic. While remote work in these companies has always been a free choice, in many other companies it is a new privilege. To regulate this privilege, the vast majority of our case companies have formulated guidelines or policies for the post-pandemic times released for piloting or rolled out in the entire company as early as in June 2021 or as late as in January 2022, with the majority being communicated in the fall 2021. These include policies, regulations, and instructions for flexibility, remote work, or hybrid work (GFT, CQSE, Tietoevry, 'GlobCo', Malvacom, Telenor), policy for work from anywhere (InterSoft), corporate rules such as rule of conduct (SB1 Utvikling) and rules of engagement (Telenor Sweden), corporate strategies such as "Future of the Storebrand", "QualityMinds Company guide", and "Sbanken post COVID", or principles for flexible/hybrid work (Kantega, Blank, 'FSwed'). The format of the new policies varies from separate contract agreements to Word documents (length varies from 4 to 18 pages), collections of slides or simply status emails or intranet posts.

**Some companies have centralised restrictions for WFH and some do not**
Slightly more than half of the companies in our dataset decided to restrict the flexibility regarding the extent of remote work. Two companies ('SpanCo' and 'KNor') introduced a special status to remote employees, who once choosing that option are expected to be remote (fully in SpanCo and 2-3 days a week in Knor). Seven companies restricted the proportion of time spent in the office vs home. GlobCo headquartered in Sweden permit remote work from home 50% of the time within a year motivated by the local regulations in Sweden, which are also applied in the company's remote site in India. Swedish Malvacom limited WFH to max 3 days/week, Sbanken, KnowIT and FSwed limited WFH to max 2 days/week, while GlobCo's Chinese site to only 1 day/week. Additionally, Sbanken specifies one mandatory office day for everybody. Finally, five companies (Storebrand, Telenor Sweden and Pakistan, SB1 Utvikling and Kantega) stated that being fully remote is not an option and that at least one day per week everyone shall spend in the office. In contrast, QualityMinds, CQSE, GFT, all sites in Tieto, InterSoft, Blank, and Telenor in Norway, Denmark and Finland do not have any centralised restrictions and provide flexible choices in their work policies.

**Flexible choices are not always fully flexible for individuals**
One may think that flexibility of choice means one can wake up in the morning and freely decide: "Today, I want to work in the office", or from home for that matter. Our analysis suggests this is not the case. In some companies, the freedom to make an ad hoc decision about where to work today is restricted by the need to align personal choices with other team members or customers. Most companies included in our study ask individuals to agree about the extent of working from home in the team (see the handshake icon in Figure 3). In Telenor Norway and Telenor Denmark, each team is explicitly expected to document their agreement in a document called a "Team Manifesto", in which the members describe which meetings, tasks or functions should preferably be conducted virtually, which can be carried out in a hybrid/mixed mode with some co-located participants and some calling in, and which require physical presence in the office. Further, team members are invited to discuss and agree on the ways to ensure that no team members get isolated or systematically left out of discussions taking place at the office, ways to build the team, keep the team-spirit and the team culture, and practices to help new hires learn from the more experienced ones. Also, the teams define which weekdays all members should come into the office for joint meetings. It is important to note that the team manifestos are living documents in which the teams make continuous updates as they learn about these new ways of work. At CQSE, the entire team is expected to manage their working mode autonomously, including vacations that team members shall take, with no formal approval process (e.g. by the team lead). Team-based decisions without management approval are also made in Telenor Norway and Sweden, Storebrand, SB1 Utvikling and Sbanken, while many other companies require a team lead, or an immediate manager to approve the team-made choices.

Besides the need to align with others, some companies make the individual WFH choices binding for a longer period of time (typically 12 months). 'InterSoft' allows employees to choose whether to work predominantly in the office (max 1 day a week from home) or predominantly from home (max 1 day a week from the office) with a binding period. Similarly, SpanCo, Malvacom, QualityMinds and 'KNor' employees who choose to work predominantly remotely can sign a special contract, which is binding for an agreed time period. These restrictions are often motivated by the local legislation (such as insurance that must cover home office), predictability of office presence, availability of a personal work desk or other office benefits (subsidised lunches, internet, or electricity).

**WFH is not for everybody**
The rules specified in the future hybrid work policies do not always apply to every single employee in the company. Many companies reserved the right to make exceptions from the WFH privilege for specific roles, jobs, or customers projects. Some companies (KNor, Knowit, Malvacom) put suitable home office equipment as a prerequisite for WFH. Some others (QualityMinds, CQSE, 'SpanCo', 'GlobCo', Telenor Denmark) can prohibit work from home if employees underperform or fail to collaborate with team members and customers. Employees who carry out technical support, customer support or depend on security-critical infrastructure in several companies may be required to be onsite, or to work in shifts. Finally, 'GlobCo' China restricts work from home for newly onboarded employees in their first half year of employment. Overall, the main difference between the companies is in how high up in the hierarchy the decisions about hybrid work policies are made. When decisions are made by the CEO or executive management team, it is hard to cater to the different needs of employees as top-level managers do not have the fine-grained information to do that. However, a beneficial aspect of organising in this way is that the rules will be the same for everybody. When decisions about hybrid policies are made on the lowest team-level in the company, it is easier to cater to individual needs of team members and their tasks. However, such an arrangement will lead to different benefits and practices across teams which might be viewed as unfair and potentially leading to unhealthy biases.

**Remote rarely means from anywhere**
If working remotely is permitted, can "remotely" mean "from anywhere"? This question is of interest for those, who have moved into new accommodations during the pandemic, often further away from the office location, as well as for expats who have moved globally to their homeland countries to spend the pandemic with the extended family members. We learned that although not always regulated in the policies, most companies participating in our research had restrictions for relocation and only allowed employees to move within the same country, primarily due to tax implications and insurance limitations (see Permanent relocation column in Figure 4). For similar reasons, Tietoevry India limits relocation to only within the same economic zone, while Norwegian Blank and Storebrand, Swedish 'GlobCo' and Telenor Sweden allow employees to move within a reasonable commute distance. Following contemporary trends, Storebrand introduced a policy to minimise business travel, which is also used as one motivation for why not to allow employees to live on a far distance.

Notably, it is not that surprising that many companies do not permit global relocation, since such transitions require awareness of the local regulations. For example, CQSE reports being required to adjust salaries to the cost of living of the employees when relocating. To "test the water", Malvacom has allowed one of remote employees to work from Spain, where he is staying an extra week after vacation. The company manager reports that if the experiment is successful, the company will consider opening up an option of broader relocation.

On the contrary, QualityMinds, CQSE, and 'InterSoft' permit global relocation, although upon prior approval. All practical questions related to relocation in these companies are supported by HR departments. 'InterSoft', which employs engineers from all over the world, restricts the choices of relocation to countries with local legal corporate presence, and reports that in 2021, 129 employees have already moved globally and 286 employees in the US alone have moved to another state. International relocation can be further restricted due to customer needs, security regulations, access restrictions to infrastructure or if it impedes efficient collaboration.

**Companies take responsibility to ensure ergonomic conditions in home offices**
In many European countries, companies are legally required to ensure an appropriate workplace for their employees. But what happens when the workplace is in someone's home? Naturally, employers do not have access to one's home to check the suitability of the work environment. In some companies, the responsibility for ensuring proper equipment and a good work environment lies with the employees who want to work remotely, while some companies condition WFH to the ability of the immediate management to check the suitability of the home office, which is made explicit in special agreements. Alternatively, some companies require employees who apply for hybrid working to declare that their home work environment is in line with the national rules and regulations and does not entail unfortunate physical strains.

Notably, many companies in our study support home office equipment beyond the necessary IT equipment. Reimbursement programs for home office equipment are active in Storebrand (500€), SB1 Utvikling, 'InterSoft', Sbanken, 'KNok', Blank (beyond budgeting), Kantega (1'000€), and CQSE, which provides full equipment either at home or in the office and which has offered 1'000€ to purchase office furniture during the pandemic. More advanced support includes covering internet fees (Blank, SB1 Utvikling, Tietoevry Norway, KNok (up to 50€), 'SpanCo' (10 €/month)), electricity (10 €/month in SpanCo), and subsidised lunch benefit (Tietoevry Finland).

**Insurance at work (or home) is also an important question**
The underlying problem with flexible working from home is that companies are responsible if something happens "at work", which is typically regulated in insurance policies. To address these challenges, some companies extended the common insurance to cover flexible work times (24/7) and employees' home addresses. In some countries such insurances are mandatory on a national level. For example, German regulations for mobile workers extended the incident insurance to cover not only working from home, but employee trips to get a lunch or drop off children in schools or daycare, and even smaller "trips" to get a drink in the kitchen or visit a restroom, all of which were previously not insured.

| Companies | Permitted options | Comments | Level of approval | Permanent relocation |
|---|---|---|---|---|
| Blank | | | | Within commute distance |
| GFT | | | | Within the country |
| Telenor Finland | | | Agreement in a work unit | Within the country |
| Telenor Norway | | | Agreement in a work unit | Within the country |
| CQSE | | | Agreement in a work unit | Globally |
| InterSoft | | | Management approval | Globally, restricted countries |
| Telenor Denmark | | | Management approval, Agreement in a work unit | |
| QualityMinds | | | Management approval, Agreement in a work unit | Globally |
| Tietoevry Czech | | | Management approval, Agreement in a work unit, Agreement with customers | Within the country |
| Tietoevry India | | | Management approval, Agreement in a work unit, Agreement with customers | Within the economic zone |
| Tietoevry Norway | | | Management approval, Agreement in a work unit, Agreement with customers | Within the country |
| Tietoevry Finland | | | Management approval, Agreement in a work unit, Agreement with customers | Within the country |
| Tietoevry Sweden | | | Management approval, Agreement in a work unit, Agreement with customers | Within the country |
| KNor | | Hybrid staff can and must work from home 2-3 days/week | Management approval | Within the country |
| SpanCo | | Choice between 100% remote or 100% from the office | Management approval | Within the country |
| Storebrand | | Fully remote is not an option | Agreement in a work unit | Within commute distance |
| Telenor Sweden | | Fully remote is not an option | Agreement in a work unit | Within commute distance |
| Telenor Pakistan | | Fully remote is not an option | Management approval, Agreement in a work unit | Within the country |
| SB1 Utvikling | | At least one common day in the office per week | Agreement in a work unit | Within the country |
| Kantega | | Fully remote is not an option | Agreement in a work unit, Agreement with customers | Within commute distance |
| GlobCo India | | 50% of the time within a calendar year | Management approval | Within commute distance |
| GlobCo Sweden | | 50% of the time within a calendar year | Management approval | Within commute distance |
| Malvacom | | Preferably no more than 3 days/week | | Within the country |
| KnowIT | | Preferably no more than 2 days/week | Management approval, Agreement in a work unit, Agreement with customers | Within the country |
| FSwed | | Preferably no more than 1-2 days/week | Management approval, Agreement in a work unit | |
| GlobCo China | | Max 1 day/week at home | Management approval | Within commute distance |
| SBanken | | Max 2 days/week at home, but not on Wednesdays | Agreement in a work unit | Within the country |

**Figure 4:** Overview of the company policies and regulations for working from home options.

Legend:
- Never at home — *All days in the office*
- Occasionally from home — *Occasionally at home; Sometimes at home; 1 day here and there at home; Parts of the day; 1-4 days/month; Less than 1 day/week*
- Less than half week — *1 day/week from home; Most days in the office but 1-2 days/week at home; 2 days / week*
- Half the time — *50:50; Equally much in the office as at home; 2-3 days/week*
- More than half week — *3 days/week; 4 days a week from home; Once a week or so in the office*
- Occasionally in the office — *1 day here and there at home; 2-3 days/month in the office; 1 day/month in the office; Almost always from home; Mostly at home, Sometimes in the office*
- Always at home — *Always at home; All days at home; 5 days/week at home; Primarily at home*
- Unsure — *Don't know*

Required approval for WFH:
- Management approval
- Agreement in a work unit
- Agreement with customers

# Concluding discussion

In this article, we illustrated the new trends in the demands for flexibility among the company employees and modifications to work policies that are emerging in response to these demands. In the following, we discuss a few important implications.

The demand for flexibility with respect to the privilege to work from home is apparent in our study. As well as the inability of the companies to satisfy all needs. This is because the needs of the employees vary greatly. The majority of the survey respondents in our study demand flexibility, similarly to related surveys from the US (Barrero et al., 2021 a) and the UK (Taneja et al., 2021) that suggest that the majority will switch to two or three days per week working from home after the pandemic. Yet, there are also the extreme groups of those who want to continue working entirely from home, and those preferring to work full time from the office. Along with personal preferences, it is fair to assume that many employees will have their preferences for others' presence or absence in the office. There will be employees who come to the office to collaborate with their teammates and colleagues, as well as recent new hires who are willing to learn from more senior colleagues, who will be upset if nobody else is present. This is why we believe that **the biggest challenge for the companies today is to find a way to accommodate the diverse needs of the employees.**

What many companies in our sample saw as the common denominator with respect to WFH is to opt for a hybrid workplace – office days mixed with WFH days, often delegating the responsibility for agreeing on the office presence to the teams or mediated by the immediate managers. While this sounds like a reasonable compromise, the mandatory office presence seems to be looked at as a discriminating policy, at least the way it is portrayed in the news. Similarly to the companies in our dataset, the big IT giant Apple emphasised the importance of in-the-office collaborations and only agreed to let employees work from home two days a week, with limited exceptions. Media reports that this decision was received with a great resistance[1]. This is a huge shift from previously well-known flexibility stigma (the negative perception towards those who work flexibly) (Chung 2018). In fact, a study looking into employee retention shows that 40% of employees who currently work from home, even if only one day a week, would seek another job if employers require a full return to the office (Barrero 2021, b), as well as most workers would accept sizable pay cuts in return for the option to work from home two or three days a week (Barrero 2021, a). Further, Barrero et al. suggest that the re-sorting of workers with respect to the scope of remote work has already started (Barrero 2021, b). Therefore, we observe that **in contrast to decades of stigmatisation of working-from-home, we might see the rise of stigmatisation of the WFH restrictions.**

But will a hybrid workplace be the winning strategy? After all, the absence of colleagues will inevitably upset at least those who prefer to work in the office. One possible development scenario is that along with the hybrid workplaces and predominantly remote companies (remote-first), there will be predominantly office-based companies (let's call them office-first). The choice is likely to differ depending on industry characteristics and the already existing corporate culture. Our analysis of the corporate work policies suggests that such choices might vary on a team level and remote-first teams and office-first teams are likely to coexist within the same company (we already see the evidence of this among our Norwegian industry partners). Yet, many companies admitted being concerned that without employees physically present in the office, the innovation and creativity, competence development, knowledge sharing, company culture and the sense of belonging are likely to suffer, as also suggested in the large study of remote work at Microsoft (Yang et al, 2021). These companies explicitly state in their policies that the office is the main place of work and even that remote work should not be seen as a right but as an appropriate way to solve or facilitate a specific situation as written, for example, in the WFH policies of Malvacom and QualityMinds. These two SMEs both focus on building strong teams in which team members feel like "family members" rather than workmates, and the daily presence in the office is considered as a prerequisite to achieve this. Similarly, the CEO of JP Jamie Dimon, the CEO of JP Morgan Chase, recently argued that extensive working from home is incompatible with the culture of his company where "hustling" and "creativity" is key. According to him, the remote working format is ill-suited for the kind of interactions characterising investment banking (Reuters, 2021)[2]. We can therefore foresee that **in the near future, we might see the rise in popularity of remote-first corporate strategies (or team practices), along with the office-first companies/teams occupying the niche of companies promoting teamwork and close collaboration as the core company/team values, opposed to remote work.**

---

[1] https://www.inc.com/minda-zetlin/apples-remote-work-policy-is-a-complete-failure-of-emotional-intelligence.html

Finally, we expect companies to change these policies and strategies over time as they accumulate experiences with hybrid work. So far, most companies' experiences with extensive working from home derive from the pandemic, when managers had no choice but to let their employees work from home. As the pandemic fades and companies experiment with working-from-home policies in situations with no social mobility restrictions, we expect they will make changes. These changes will be made through a dialogue centred on the "psychological contract" between employees and employers.

At the same time, it is important to note that past decisions on working from home policies are likely to shape future ones. As we have emphasised, there has been a shift in the stigma associated with working from home, where the manager trying to restrict working from home days faces more scorn, than employees expressing a wish to work from home. Employees increasingly view working from home as an entitlement, and companies that decided relatively liberal policies with high freedom to choose are likely to face strong employee resistance if they choose to rein in this freedom.

# Acknowledgement

We are thankful for all the companies participating in our study for their engagement and interest in the outcome of our research. This research is funded by the Swedish Knowledge Foundation within the ScaleWise project (KK-Hög grant 2019/0087), the S.E.R.T. project (research profile grant 2018/010), SHADE project (KK-Hög grant 2017/0176), and the Research Council of Norway through the 10xTeams project (grant 309344) and the A-team project (grant 267704).

# References

[1] Barrero JM, Bloom N and Davis SJ, 2021 (a) Why working from home will stick (No. w28731). National Bureau of Economic Research.
[2] Ford D, Storey MA, Zimmermann T, Bird C, Jaffe S, Maddila C, Butler JL, Houck B and Nagappan N, 2020. A tale of two cities: Software developers working from home during the covid-19 pandemic. arXiv preprint arXiv:2008.11147.
[3] Ralph P, Baltes S, Adisaputri G, Torkar R, Kovalenko V, Kalinowski M, Novielli N, Yoo S, Devroey X, Tan X and Zhou M, 2020. Pandemic programming. Empirical Software Engineering, 25(6), pp.4927-4961.
[4] Barrero JM, Bloom N and Davis SJ, 2021 (b). Let Me Work From Home, or I Will Find Another Job. University of Chicago, Becker Friedman Institute for Economics Working Paper, (2021-87).

---

[2] https://www.reuters.com/article/us-jp-morgan-ceo-idUSKBN2CL1HQ

[5] Nicks L, Gesiarz F, Hardy T and Burd H, 2021. How many days should we work from home? What works to improve gender equality. Research report. The Behavioural Insights Team, Government Equalities office.
[6] Taneja S, Mizen P and Bloom N, 2021. "Working from home is revolutionising the UK Labour market," VoxEU CEPR Policy Portal, 15 March.
[7] Yang L, Holtz D, Jaffe S, Suri S, Sinha S, Weston J, Joyce C, Shah N, Sherman K, Hecht B and Teevan J, 2021. The effects of remote work on collaboration among information workers. Nature human behaviour, pp.1-12.
[8] Clear T, 2021. THINKING ISSUES Loosening ties: permanently virtual teams and the melting iceberg of relationship. ACM Inroads, 12(3), pp.6-8.
[9] Smite D, Tkalich A, Moe NB, Papatheocharous E, Klotins E, and Buvik MP. Changes in perceived productivity of software engineers during COVID-19 pandemic: The voice of evidence. Journal of Systems and Software. 2022, Apr 1; 186: 111197.
[10] Smite D, Mikalsen M, Moe NB, Stray V, and Klotins E 2021, From Collaboration to Solitude and Back: Remote Pair Programming During COVID-19. In International Conference on Agile Software Development, pp. 3-18. Springer, Cham
[11] Kane GC, Nanda R, Phillips A, and Copulsky J, 2021. Redesigning the Post-Pandemic Workplace. MIT Sloan Management Review 62(3), pp. 12-14.
[12] Choudury P, 2020. Our Work-from-Anywhere Future. Best practices for all-remote organizations. Harvard Business Review 98(6): 58-67.
[13] Gensler Research Institute, 2021. US Work From Home Survey. https://www.gensler.com/workplace-surveys/us-work-from-home-survey/2020
[14] Chung H, 2018, Gender, flexibility stigma, and the perceived negative consequences of flexible working in the UK. Social Indicators Research. https://doi.org/10.1007/s11205-018-2036-7.
[15] Pratt JH. 1984. Home Teleworking: A Study of its Pioneers. Technological Forecasting and Social Change, 25(1): 1-14.
[16] Allen M. 2017. The SAGE Encyclopedia to Communication Research. SAGE Publications INC.
[17] Nguyen MH, Armoogum J. 2021. Perception and preference for home-based telework in the covid-19 era: A gender-based analysis in Hanoi, Vietnam. Sustainability, 13 (6).
[18] Menard, Scott. (2002) *Applied logistic regression analysis*. Vol. 106. Sage, London.